# Advances in electron backscatter diffraction

Alex Foden, Alessandro Previero, and Thomas Benjamin Britton*

Imperial College London
*b.britton@imperial.ac.uk

**Abstract**. We present a few recent developments in the field of electron backscatter diffraction (EBSD). We highlight how open source algorithms and open data formats can be used to rapidly to develop microstructural insight of materials. We include use of AstroEBSD for single pixel based EBSD mapping and conventional orientation mapping; followed by an unsupervised machine learning approach using principal component analysis and multivariate statistics combined with a refined template matching method to rapidly index orientation data with high precision. Next, we compare a diffraction pattern captured using direct electron detector with a dynamical simulation and project this to create a high quality experimental 'reference diffraction sphere'. Finally, we classify phases using supervised machine learning with transfer learning and a convolutional neural network.

## 1. Introduction

Electron backscatter diffraction (EBSD) is a commonly used microscopy technique [1, 2] as it provides rich microstructural maps. In brief, each diffraction pattern contains information about the scattering and subsequent diffraction of the electron beam with respect to the sample and how these scattered electrons interact with the 2D electron sensitive detector. This detector is usually a phosphor screen optically coupled with lens to a charged coupled device (CCD), but new detectors include phosphor-fiber-CMOS, phosphor-lens-CMOS and direct electron detectors [3-5]. Once the diffraction patterns are captured, they are stored and can be interrogated to provide rich information about the material.

Conventional analysis uses image processing, including background correction and band localization within the Hough/Radon transform of the pattern [6-8]. Indexing of the bands is performed using a look-up-table of the expected interplanar angles and then the crystal orientation can be determined [5], and recently that and open source solution has been made available [9]. Recent developments in EBSD analysis utilize cross correlation approaches, originally applied in the High (Angular) Resolution EBSD (HR-EBSD) method introduced by Wilkinson et al. [10, 11] (with later key updates [12-15]).

Early in the HR-EBSD analysis, for one early example, we wanted to analyze the GND density within regions of common orientation in rolled titanium, so-called "macrozones" [16] and we ran the map with multiple 'reference points' to capture the angular deviations of unsolved regions. The map was reconstructed by selecting the GND density obtained from the reference with the highest cross correlation peak height and this is a crude form of 'template matching'. Template matching is a common computer vision strategy for matching two or more similar signals, where we compare a test signal against a library of potential templates. The best fitting match (i.e. the one which the highest similarity index) is taken to best represent the unknown signal and can be used to label the test image. At the time of the macrozone study, comparison of experimental patterns with simulations was being attempted specifically to try to access the absolution elastic strain, but this remains incredibly difficult as there several issues that are still yet to be overcome [17]:

More recent template matching methods have appeared now that we have more ready access to high quality dynamical simulations. The so-called Dictionary Indexing (DI) method [18-21] uses simulations as templates for orientation and phase labelling. The best normalized dot product (subtly different to the cross-correlation coefficient) is selected for the best match. Variations of the whole pattern matching approaches have begun to develop. Notably Wilkinson et al. [22] used unsupervised machine learning, via principal component analysis and multivariant statistics, to reduce the experimental data set to a few patterns and thus reduce the number of cross correlation operations that need computing. Furthermore, Foden et al. [23] have developed a FFT-based cross correlation which includes an iterative interpolation step, to reduce the compute costs of template matching.

As an interesting alternative approach to template matching with large numbers of templates, we can reconsider this challenge and explore cross correlation on the sphere. This builds upon much earlier work of Day [24], who (through reprojection) remapped experimental patterns on the sphere. Recently Hielscher et al. [25] used the non equi-spaced FFTs to perform cross correlation directly on the sphere. Furthermore, Winkelmann et al. [26] extended this idea using symmetry operations to directly produce an experimental-based reference template, which is very interesting as it provides us a route to access information about materials where simulations may be challenging (e.g. in minerals with varying composition) and/or materials with interesting symmetry (e.g. the quasicrystal example demonstrated by Winkelmann et al. [26]).

Improving signal to noise, beyond counting for longer, requires investment in hardware. An intermediate approach to improve detection is to use more direct coupling of the phosphor and photosensitive device, e.g. through fiber optic coupling with oil-to-glass interfaces. However, these indirect systems will be inefficient as fundamentally an electron signal is being convert to light and then back to electrons, rather than simple counting of the total number of electrons in a 2D grid. Improvement can also be achieved with direct electron detectors (DED) [27]. Wilkinson et al. [5] demonstrated the potential of DED when a CMOS device was used for direct capture of silicon diffraction patterns. Subsequently, the Medipix and Timepix hybrid pixel detectors. Vespucci et al have demonstrated their effective use as EBSD detectors with the presentation of crisp EBSD patterns of semiconductor materials [3, 4, 28-31]. With these higher quality patterns, we can now ask if there are interesting new ways to explore them, potentially with machine learning.

In this manuscript, we will provide a few case studies to highlight the benefits of these recent advances in EBSD detection and analysis and show promise for further advances.

## 2. Open Source EBSD

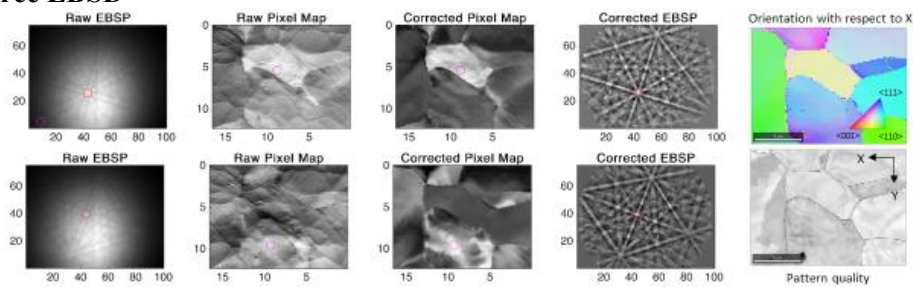

**Figure 1**: Microstructural imaging using AstroEBSD on a data set from deformed interstitial free steel. Diffraction vectors for the maps are highlighted in red. The raw pixel map reveals surface topography, while the background correction focusses on channelling out, and thus highlights internal microstructure. Diffraction vectors for the pixel maps are highlighted in red within the EBSPs. Orientation map and pattern quality are generated with AstroEBSD indexing. The pixel map locations are highlighted in purple.. [The units of the EBSPs are in pixels, with 4x4 binning; and the units of the maps are in μm].

Many EBSD hardware manufacturers now enable saving of data sets via easy to use the Hierarchal Data Format v5 (HDF5®) [19] which is well suited for software development packages like Matlab and Python. We use a small data set (https://doi.org/10.5281/zenodo.2609220) taken from plastically deformed interstitial free steel and analysis it with AstroEBSD [6]. This software tool includes a series of typical routines for routine data handling of EBSD patterns and associated information, including background correction, and indexes the patterns. In Figure 1 for two electron backscatter patterns (EBSPs) we perform background correction. Next, similar to Wright et al. [20], we perform direct imaging with our EBSD detector (Figure 1) with one single diffraction vector (pixel) for each map and we select two diffraction vectors located at bright zone axes. In the 'raw' signal image (i.e. before background correction) we see the combined effects of 'channeling-in' as well as 'channeling-out' and topographic contribution. In the background corrected based map, we now observe subtle variations in the grain orientation (and lattice strain) which change the intensity of features within the deformed grain. While this example shows largely that 'pretty pictures' can be generated, it is also easy to imagine that selection of an appropriate diffraction vector or a diffraction based virtual aperture could provide insight into (for example) lattice rotations of a particular type or imaging based upon the presence (or absence) of a specific electron channeling condition due to the precipitation of a superlattice phase.

Furthermore, we can use AstroEBSD to index the patterns and map them to orientation space to create rich microstructure-based maps. In practice, the outputs are very similar to those which are obtained from a commercial system but the opportunity here is that we now know exactly what is being computed (and thus can understand its limitations as well as benefits). Furthermore, we can load the AstroEBSD data directly into MTEX [21] and manipulate the orientation data as required

## 3. Templated matching with refinement and principal component analysis

Next, we demonstrate refined template matching, and exploit this method to improve the indexing of the iron data set which is initially indexed using the principal component analysis method.

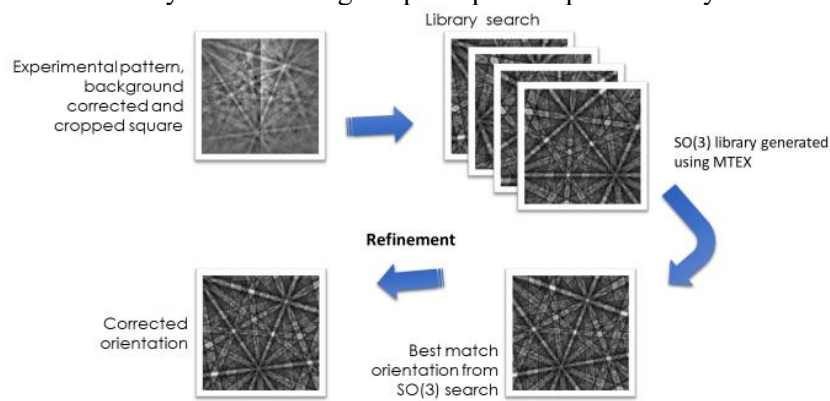

**Figure 2**: Refined template matching approach, adapted from Foden et al.[23]

The refined template matching scheme is presented in Figure 2 (figure adapted from Foden et al. [23]). Running this on the deformed iron data set is relatively quick (less than 1 hour), as there are only 9130 patterns this map is small, and we might want to consider methods to reduce the number of correlations required. Therefore, we propose optimization of the refined template matching approach when combining this with the principal component analysis (PCA) and multivariate statistics approach presented by Wilkinson et al. [22].

In brief, the PCA enables us to reduce the $P \times Q$ diffraction patterns (components) that span $M \times N$ points (variables), down to a limited number of common components (i.e. ~1 pattern per grain). These are put into a data array, and normalized, before they are decomposed into their principal components. Then the components are rotated to create patterns that can be pattern matched (see Figure 3). At this stage, the orientation of each labelled region is uniform, but representative (this represents the method of Wilkinson et al. [22]). To optimize this further, we can use the second stage of our refined template matching method [23] to update the orientation of each diffraction pattern from every point in our map based upon a refinement step, optimized to start with the label obtained from the PCA-MSA step. The results of this optimization is shown within the refined labels map of Figure 3, where the smoother orientation gradients are recovered within and between each label obtained from the loading map.

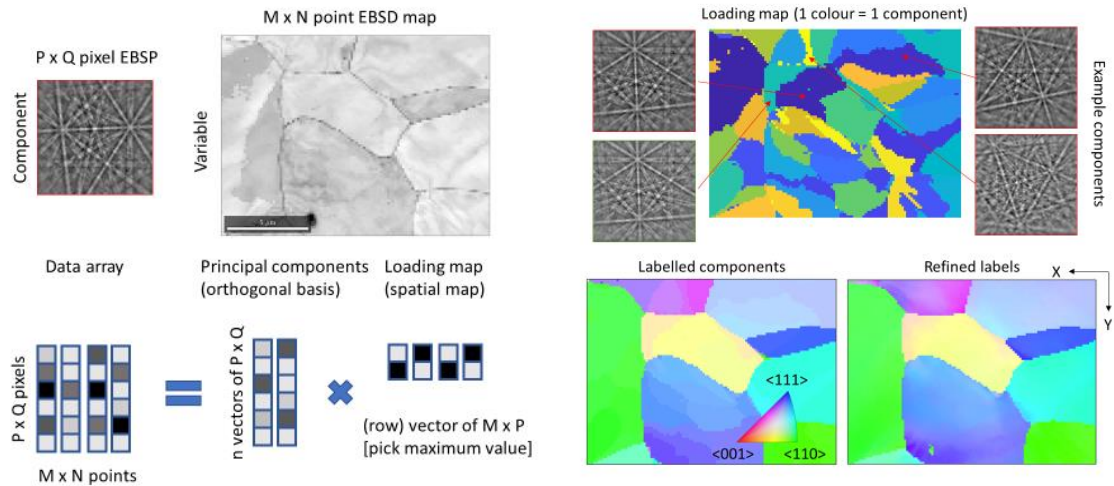

**Figure 3**: Schematic and application of principal component analysis, followed by pattern matching to measure the orientation of each label, and then use of the refined templated matching for the labels.

## 4. Spherical EBSD with Direct Electron Detection Patterns

If we know the pattern centre (closest point to the screen and detector distance), we can use simple geometry [32] to reconstruct a portion of the diffracting sphere. We collect a silicon diffraction pattern and capture a pattern with a DED (ADVACAM MiniPIX which contains a 256 x 256 pixel Timepix 2 detector with a 55 μm pixel size). The experimental pattern was background corrected using AstroEBSD routine and matched with a simulated dynamical diffraction pattern [33] (Figure 4).

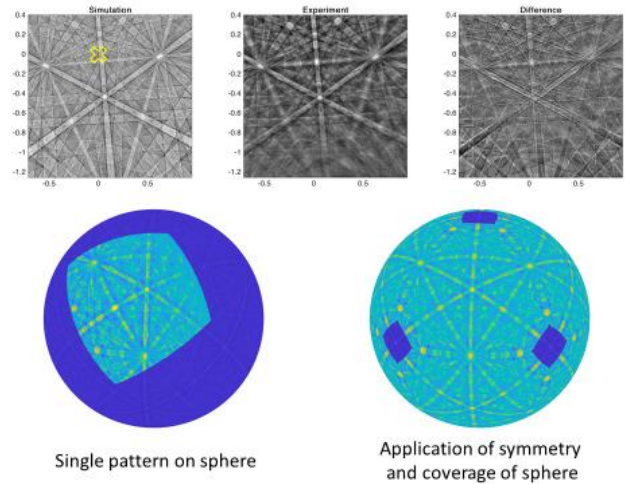

**Figure** 4: Simulated and experimental patterns of silicon, illuminated with a 20 keV electron beam. The top line shows the patterns, and their difference, as represented in the gnomonic frame of reference. The pattern centre is highlighted with a yellow cross. The units of these patterns are in the gnomonic coordinate system. The bottom line shows them reprojected onto the sphere.

Projection of one pattern onto the sphere shows that we can have sensible alignment of the band centres with the great circles of the sphere (Figure 4). We can develop this projection further and use the symmetry operators of the silicon crystal (i.e. 24, as it has cubic symmetry) to rotate the pattern on the sphere and overlap them (this mathematics is used in the spherical EBSD indexing paper by Hielscher et al. [25]). In Figure 4 for overlapping pixels, the intensity can be normalised by the number of symmetrically equivalent patterns which are overlapped. This pattern covers more of the sphere and due to the pattern centre used here (with a long camera distance) and crystal orientation we do not have complete coverage of all the diffraction vectors, leaving absence of information at the <100> zone axes. Furthermore, variation in the contrast with respect to position of the detector results in (inadvertent) blurring of the spherical EBSD pattern.

## 5. Machine Learning and EBSD

Machine learning is a broad description of algorithms that generate sophisticated statistical models of input data for fast or unusual analysis. They include supervised methods such as convolutional neural network (CNN) models that create a series of regressions, using weightings and filters, which can be used to classify similar types of data. The CNN is trained to find areas of similarity and contrast as to fit the training data and subsequently classify test data. While the application of machine learning and EBSD is not new (neural networks were used in 2003 by Schwarzer and Sukkau [34] to verify automated evaluation of crystal orientations) improved development of machine learning tools (e.g. through toolboxes in Matlab and Python packages) combined with faster computers with more memory means that we can re-explore their utility for EBSD analysis.

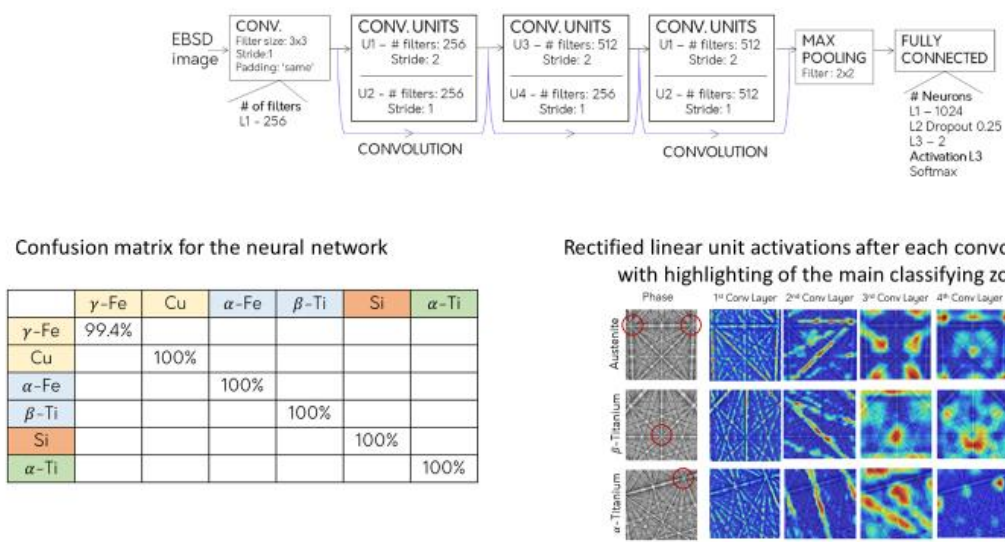

**Figure 5**: Application of machine learning for phase classification for EBSD patterns

We focus this section on the application of machine learning for phase classification. To demonstrate potential, we will explore using dynamical simulations alone. We start with generation of a series of dynamically simulated patterns from 8 potential phases: $\gamma$-iron (FCC), copper, $\alpha$-iron (BCC), $\beta$-titanium (BCC), silicon (diamond cubic), and $\alpha$-titanium (HCP). These patterns represent similar crystal structures and diffraction patterns and combinations of phases that may be easily confused (e.g. FCC and diamond cubic) by conventional analyses. We generate a library of templates from the master dynamical simulations created with Bruker DynamicS, as projected into a gnomonic frame of reference for a typical camera geometry.

Next we create a convolutional neural network using transfer learning to take a pre-trained network, using AlexNet (a straight convolutional neural network developed by Alex Krizhevsky, originally for the ImageNet competition [35]). We train this to classify each phase using multiple convolutions and filtering steps (see Figure 5), with learning rates were optimised for our classification challenge (with rates of 3, 6, 9, 18 and 24 employed for the six-phase classification). The corresponding confusion matrix for our optimised network is shown in Figure 5 and shows extremely impressive results that highlight potential to distinguish very similar phases apart.

From the CNN, we can also plot the activations of each layer of the network as applied to an input signal to 'understand' how the neural network 'sees' contrast between the classifiers and assigns each test pattern to the trained classifications. We plot the rectified linear unit activations after each convolutional layer (Figure 5). These weightings highlight that the first layer is identifying edges, like the Burn's algorithm; the second layer is identifying major bands, like the conventional Hough/Radon analysis; and increasing layers tend to explore the structure of the zone axis for classification. These activation maps provide insight into how we might develop deterministic approaches to classify very similar diffraction patterns, such as the diamond cubic silicon and the two face centred cubic phases ($\gamma$-Fe and Cu shown here).

## 6. Discussion

It is remarkable that we continue to see advances with EBSD, considering the technique was commercially developed almost >25 years ago. In the present article we have not discussed 3D EBSD, nor transmission Kikuchi diffraction [36], nor the application of EBSD for *in situ* and *in operando* studies [37] (e.g. with heating or deformation). Instead we have focussed on technique developments that explore how software can enable us to improve the quality and accessibility of EBSD derived insight into materials and combined this with a look at how new hardware such as direct electron detectors may help us further.

The AstroEBSD example for point-based imaging may seem trivial, and yet we can already see how the 4D scanning transmission electron microscopy (STEM) community is being transformed by the use of virtual apertures [38]. We could imagine the generation of apertures that track with an update to the underlying crystal orientation (e.g. link HR-EBSD with virtual aperture based imaging) to explore more subtle variations such as the change in local structure and/or effective scattering. This is like using STEM imaging to suppress the presence of bend contours in TEM except that in EBSD experiments we can move our virtual aperture (e.g. using HR-EBSD to map the same imaging point) and reprocess our data as many times as required to optimise contrast.

The Open Source software toolbox that adheres to well described orientation conventions [9, 32] and can be inspected and tested by the user and this increases confidence and promotes easier development, such as the sharing of routines to make PCA-MVS method and spherical remapping almost trivial to perform. This is greatly facilitated through open data formats such as HDF5. The PCA-MSA, with refinement, method shows that we can link techniques together to provide computationally efficient and statistically reliable (to amplify signal to noise) processing strategies when needed. Notably this results in a better measurement of the orientations near grain boundaries.

The DED pattern demonstrates high quality image capture, despite the low number of pixels, the quality (MTF and DQE) is very high. High signal to noise in these patterns means we are likely to see advances in EBSD of beam sensitive materials with short exposure times [39]. Unfortunately, with the increase in quality we see scattering path effects (coherent vs incoherent sources, voltage variations, band contrast reversal) which affect the contrast and may affect precise measurements e.g. a whole pattern based cross correlation routines [40-42], as these measurements may run into systematic issues when applied 'in anger' to higher quality experimental patterns. Furthermore, we may also be limited if less likely assumptions about the scattering physics are assumed [43].

Finally, the implementation of machine learning may provide us with 'blackbox' solutions that will be useful [44]. Machine learning as applied to diffraction pattern-based image problems [45] and EBSD [34] is not new and yet due to new access to techniques and richer datasets we are seeing an exciting resurgence [46, 47]. These may include the classification of similar phases, provided the algorithm is suitably trained and optimised with a result in mind (i.e. the problem is suitably constrained and the blackbox is not overfitted nor used to perform ill conditioned extrapolation). If an appropriate machine learning strategy is employed, we may discover new approaches that seed deterministic insight into diffraction pattern analysis, such as detection of specific features or structures that truly represent high contrast between two very similar phases.

# 7. Conclusion

We demonstrate that EBSD can advanced using:

i. Open source algorithms that provide transparent manipulation of diffraction pattern data.
ii. Combinations of statistical methods, brute force pattern matching, and interpolation schemes.
iii. Use of direct electron detectors to improve the quality of the detected diffraction patterns.
iv. Exploration of machine learning approaches to provide new methods of diffraction pattern analysis.

These brief examples highlight that we continue to develop this electron crystallography technique and to apply it to interesting materials science, geomaterials, and engineering challenges to reveal new insight and understanding. The benefit of a wide capture angle, good spatial resolution, and relative access of the technique means that EBSD remains useful and prominent in our materials characterisation toolbox.

# 8. Acknowledgements

We thank: Dr Aimo Winkelmann, Prof Angus Wilkinson, Dr Vivian Tong, and Dr Ralf Hielscher for insightful comments and discussions and code snippets. We thank Ben Wood for assistance in manufacturing of the new direct electron detector system. TBB thanks the Royal Academy of Engineering for his Research Fellowship. TBB and AF thank EPSRC for DTP funding.